\documentclass[aps,prl,twocolumn,showpacs,groupedaddress]{revtex4}
\usepackage{graphicx}
\usepackage{bm}
\begin{document}

\title{Induction of phase-slip lines in a thin, wide superconducting film by an rf
electromagnetic field}

\author{V.~I.~Kuznetsov}
\email[Electronic address:]{kvi@ipmt-hpm.ac.ru}
\author{V.~A.~Tulin}
\affiliation{Institute of Microelectronics Technology and High Purity Materials,
Russian Academy of Sciences, 142432 Chernogolovka, Moskow Region, Russia}

\date{\today}

\begin{abstract}
The appearance of phase-slip lines, induced by an rf field, was observed
experimentally in wide superconducting films in which the destruction of
superconductivity by a dc current is associated with the formation of phase-slip
lines. The characteristics of the separation of the film into superconducting and
nonuniform, isothermal, nonequilibrium regions (phase-slip lines) under
electromagnetic irradiation were studied.
\end{abstract}

\pacs{74.40.+k, 74.25.Qt, 74.78.Db, 74.78.-w, 74.50.+r}

\maketitle

1. When a current exceeding the critical current ($I > I_{c}$)
flows through a thin superconducting film at temperatures close to
the critical temperature ($T < T_{c}$), phase separation of the
film into superconducting and spatially localized, nonequilibrium
regions is possible. In narrow films whose width is less than the
coherence length [$w < \xi(T)$], these regions, in which the
modulus $\vert \Delta \vert$ of the order parameter of the
superconductor periodically vanishes and the phase of the order
parameter changes by $2\pi n$ ($n$ is an integer), are called
phase-slip centers (PSC) \cite{Ivlev}. At temperatures close to
$T_{c}$ and with good heat transfer, wide films [$w > \xi(T)$] can
form nearly isothermal, nonequilibrium, nonuniform regions
(phase-slip lines), whose length 2 $l_{E}$ is equal to twice the
penetration depth of a longitudinal electric field into the
superconductor.

The appearance of phase-slip centers has been studied extensively
\cite{Ivlev}. The main experimental studies of phase-slip lines
(PSL) are presented in Ref. \cite{Vol}. We know of only one
theoretical work on the formation of PSL during the passage of a
dc current through a film \cite{Lem}. The discovery of "nitrogen"
superconductors has renewed interest in the development of devices
that operate on the basis of the switching from the
superconducting state to the resistive state. The formation of PSL
is a possible realization of the resistive state. In wide films
there exist, besides PSL, magnetic flux lines. The motion of flux
lines can influence the dynamics of PSL and the PSL themselves can
influence the dynamics of the flux lines \cite{Siv}. The
experimental investigations of PSL performed thus far do not
completely explain the physics of such lines. It follows from
Refs. \cite{Vol, Ilich, Kuz} that PSL in wide films are similar to
PSC in narrow films. The I-V characteristics of samples containing
PSL have a step character and consist of a series of linear
sections, whose resistance is a multiple of the resistance of a
single PSL, i.e., $R\simeq nR_{0}$. The resistance to current flow
of a single PSL (just as a PSC) is $R_{0}=2\rho_{n}l_{E}/wd$
\cite{Ivlev, Vol}, where $\rho_{n}$, is the normal-state
resistivity of the film, and $d$ is the thickness of the film.

The penetration depth of an electric field is determined by
different mechanisms of the relaxation of the unbalance of the
populations of the electron- and hole-like branches of the
quasiparticle spectrum of the superconductor. In theoretical
studies these mechanisms for PSC were investigated in, for
example, Refs. \cite{Bez, Art}. In Refs. \cite{Ilich, Kuz} the
cases in which both branches of the spectrum are mixed as a result
of inelastic electron-phonon collisions and the so-called elastic
mechanism, which is the dominant mechanism when the condensate
current is sufficiently strong, were studied experimentally in
wide films. The investigation of the temperature dependence of the
PSL resistance makes it possible to determine which mechanism of
hole-electron conversion leads to the appearance of a nonuniform
electric field in the superconductor plays the main role. The
dynamics of PSL in wide films irradiated with high-frequency
radiation has still not been studied experimentally. Our objective
was to conduct such a study.

2. The samples consisted of tin films of width $w = 70$ $\mu$m,
thickness $d\approx 1000$ \AA, and length $L = 2$ mm. The films
were thermally deposited on polished silicon substrates. A sample
was placed in an 8-mm waveguide. The plane of the substrate was
parallel to the electric component of the 30 GHz electromagnetic
field. The dc I-V characteristics of the sample were recorded at
temperatures close to the critical temperature with different
irradiation power levels. We note that in tuning the radiation
generator the maximum output power $P_{max}$ was changed in each
new series of measurements. We calculated the resistance of a PSL
according to the slope of the first visible (sometimes second and
third) linear section of the I-V characteristic which was close to
the critical current.

3. To determine the mechanism of the penetration of a nonuniform
longitudinal field into a superconducting film, we investigated
the temperature dependence of the resistance of the first
phase-slip line with no irradiation. This dependence was
nonmonotonic, just as in \cite{Ilich} for phase-slip lines and in
\cite{Kadin} for phase-slip centers. In the isothermal region, the
resistance was determined by the penetration depth of the electric
field
\begin{equation}
l_{E}=\sqrt{D\tau_{Q}}=\sqrt{D\tau_{\varepsilon}4kT/
\pi\vert\Delta\vert}\;, \label{1}
\end{equation}
where $D$ is the diffusion coefficient, $\tau_{Q}$ is the
relaxation time of the asymmetry of the populations of the
branches of the quasiparticle spectrum, and $\tau_{\varepsilon}$
is the inelastic electron-phonon scattering time. Then near
$T_{C}$
\begin{equation} l_{E}=g(D\tau_{\varepsilon})^{1/2}(1-t)^{-1/4}\;,
\label{2}
\end{equation}
where $t = T/T_{c}$ and $g\approx 1$. For our samples with a mean
free path $l\approx 300$ \AA, $T_{c}\approx 3.91$ K,
$\tau_{\varepsilon} \approx 3\times 10^{-10}$ s, and
$\rho_{n}l\approx 1.6 \times 10^{-11}$ $\Omega$ cm$^{2}$
\cite{Ilich}, the temperature dependence of $l_{E}$, determined
from $R_{0}(T)$, was close to that presented above. For our films,
the penetration of the longitudinal electric field in the
isothermal region is therefore associated mainly with the
inelastic electron-phonon mechanism of balancing of the
populations of the electron-like and hole-like branches.
\begin{figure}
\includegraphics [width = 1.0\linewidth]{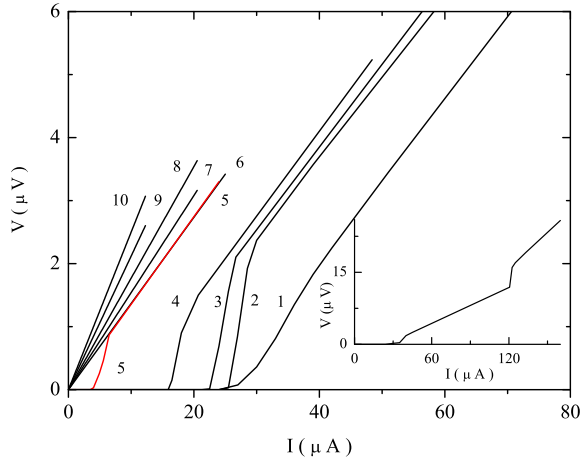}
\caption{\label{fig. 1} Initial sections of the I-V characteristic
with irradiation power in the range of suppression of the critical
current and increase in the resistance of the first linear section
from $R_{0}$ to $\approx 2R_{0}$. Curve 1: -70 dB, 2: -24 dB, 3:
-23 dB, 4: -20 dB, 5: -19.3 dB, 6: -19.1 dB, 7: -17.3 dB, 8: -16.4
dB, 9: -15.4 dB, 10: -13.9 dB (damping in decibels). $T = 3.89$ K,
$P_{c}/P_{max} = -19.1$ dB, $\omega/2\pi = 30$ GHz. Insert: I-V
curve of a sample with no irradiation, including two PSL steps; $T
= 3.87$ K.}
\end{figure}
\begin{figure}
\includegraphics [width = 1.0\linewidth]{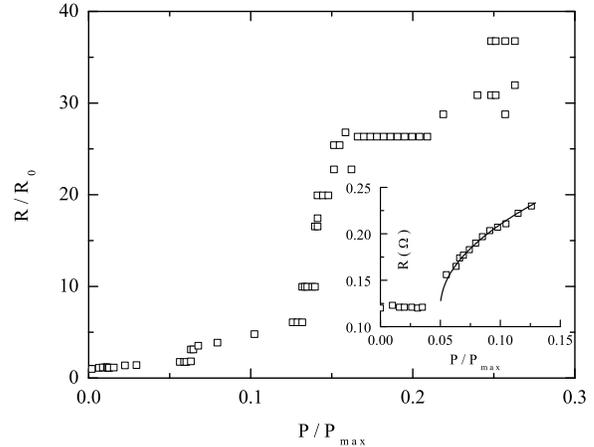}
\caption{\label{fig. 2} Normalized resistance of the first visible
PSL step near $I = 0$ as a function of the relative irradiation
power. $T\approx 3.88$ K, $R_{0}\approx 0.12$ $\Omega$,
$P_{c}/P_{max}\approx 0.01$. Inset: Resistance of the film as a
function of the relative irradiation power for $R$ from $R_{0}$ to
$\approx 2R_{0}$. $T\approx 3.88$ K, $R_{0}\approx 0.12$ $\Omega$,
$P_{c}/P_{max}\approx 0.05$. Solid line:
$R=R_{0}+b(P/P_{max}-P_{c}/P_{max})^{1/2}$, where $b\approx 0.4$
$\Omega$. This power dependence was measured in the new series of
measurements, so that the value of $P_{max}$ is different.}
\end{figure}

4. Similar results were obtained for different films in all series
of measurements. For this reason, only the characteristic and most
detailed dependences from different series are shown in the
figures. Figure 1 shows the I-V characteristic of one sample as a
function of the irradiation power at a temperature close to
$T_{c}$. The I-V characteristic of the unirradiated sample,
including two steps, is shown in the inset in Fig. 1. As the power
increases, at first the critical current can increase. This could
be associated with the stimulation of superconductivity \cite{El}.
It is not represented in Fig. 1. At some power ($\approx - 24$ dB
) the critical current is appreciably suppressed and vanishes in
an interval of several dB. We call the power $P_{c}$ at which
$I_{c}=0$ the critical power. The resistance $R_{0}$ remains
virtually constant at a power lower than $P_{c}$. The initial
section of the I-V characteristic for $P
> P_{c}$ is linear and there is virtually no hysteresis in the I-V
characteristic recorded in both directions. Despite the absence of
a critical current, the step structure of the I-V characteristic
remains up to the irradiation power at which the resistance of the
film is equal to the normal-state resistance of  the whole film.
Similar I-V characteristics were also observed with no stimulation
of superconductivity.

5. Figure 2 shows the normalized resistance $R/R_{0}$ as a function of the
relative irradiation power $P/P_{max}$, where $R$ is the resistance of the first,
linear section, which is close to the critical current, of the I-V
characteristic. The resistance $R_{0}$ is the resistance of a single PSL with
no irradiation. In the case of no critical current the resistance was
determined near a current close to zero. The measurements were performed in
the temperature range where dc overheating phenomena are weak, i.e., in the
isothermal region. As the power increases, the normalized resistance
increases, mainly by a jump, and after the jump it equals an integer. The
small nonuniqueness of the dependence is explained by the fact that in
repeated measurements there was a scatter in the powers at which a jump onto
the $n$th step occurred. Moreover, at some powers the probabilities of states
with the $n$ and ($n + m$) PSL, where $n$ and $m$ are integers, were close.
The sample jumped from one state into another. Separate points corresponding to
the second or third visible linear step are also shown in Fig. 2. In some power
intervals the resistance of the first, close to zero, step increased continuously, and
then made a more or less distinct jump to a state with $n$ PSL. We constructed
the experimental curve of the resistance versus $P/P_{max}$ in the interval
where the resistance increases continuously from $R_{0}$ up to values $\leq 2R_{0}$
for one of the samples (Fig. 2, inset). These data can be approximated by a root function
\begin{equation} R=b(P/P_{max}-P_{c}/P_{max})^{k}+R_{0}\;,
\end{equation}
where $k\approx 0.5$, and $b$ is a constant. Such dependences were also obtained
for other films.

6. A phase separation into superconducting and nonequilibrium regions (PSL),
which is induced by high-frequency irradiation, has thus been observed for
the first time in wide superconducting films. The "high-frequency"
separation is similar to the "current" separation. The possibility of
"high-frequency" separation when superconductivity is stimulated in narrow
channels was described in Ref. \cite{Iv}. To clarify the physical picture of the
process and to compare it with the theory of Ref. \cite{Iv} and the experiment of
Ref. \cite{Dmitr} for narrow films, where PSC are formed, we constructed the
temperature dependence of the relative critical power $P_{c}/P_{max}(T)$. Near
$T_{c}$ this dependence has the form $P_{c}/P_{max}\sim (1-T/T_{c})^{3/2}$ (Fig.
3). However, it is close to a linear dependence at temperatures farther away
from the critical temperature and for somewhat higher powers. In this case
the isothermal state could be disrupted. The temperature dependence
$P_{c}/P_{max}\sim(1-T/T_{c})^{3/2}$ agrees with the following experimental
results: 1) The critical current $I_{c}(0)$ with no irradiation and the
critical power $P_{c}/P_{max}$ have the same temperature dependence near
$T_{c}$ and 2) the critical current $I_{c}(P/P_{max})$ with irradiation
is roughly a linear function of the power near $T_{c}$ and $P$ close to $P_{c}$.

At a temperature very close to $T_{c}$ and low power, however, this
dependence had the form $I_{c}(P/P_{max})=I_{c}(0)-c(P/P_{max})^{1/2}$,
where c is a constant. In Ref. \cite{Kulik} the
power dependence $I_{c}(P/P_{max})$ at low power was close to
quadratic. Instabilities of different nature which result in the destruction
of superconductivity have a different temperature dependence of the critical
power. In Ref. \cite{Iv} the relative critical power of the high-frequency
separation of a film under conditions of stimulation of superconductivity in
narrow films is proportional to $1-T/T_{c}$. In the case of an instability as a
result of the pair-breaking current $P_{c}\sim(1-T/T_{c})^{3}$, for flux-line motion
$P_{c}\sim (1-T/T_{c})^{2}$, and in a strong parallel field
$P_{c}\sim 1-T/T_{c}$ \cite{Tink}. In Ref. \cite{Kulik}
$P_{c}\sim 1-T/T_{c}$ for a thin film with a transport
current and microwave irradiation, disregarding transparency. Very close to
$T_{c}$ other temperature dependences of $P_{c}$ can be expected because of the
nonequilibrium nature of the gap. The theories of Refs. \cite{Iv, Kulik} are
concerned with uniform films which transform at the critical irradiation
power $P_{c}$ into the normal state \cite{Kulik} or into a nonuniform
state with phase separation into PSC, the nonequilibrium regions immediately
filling the sample along its entire length with a definite period \cite{Iv}.
In our case, just as in Ref. \cite{Dmitr}, the number of PSL which are
formed on the nonuniformities of the film increases systematically with
increasing irradiation power above $P_{c}$. However, the temperature
dependence $P_{c}(T)$ is different from the dependences presented in
Refs. \cite{Iv, Dmitr, Kulik, Tink}.
\begin{figure}
\includegraphics [width = 1.0\linewidth]{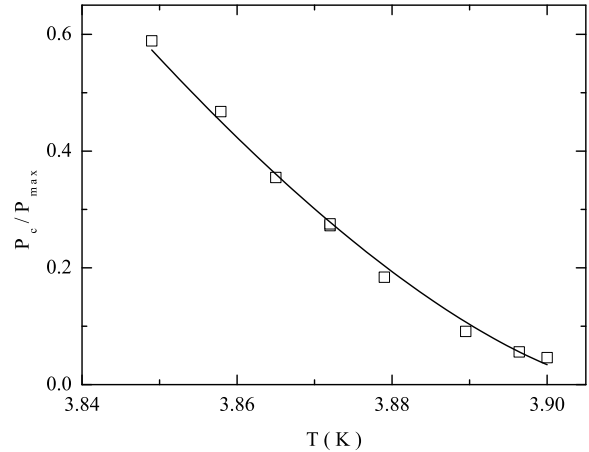}
\caption{ \label{fig 3.} Temperature dependence of the relative
critical power at temperatures close to the critical temperature;
$T_{c}\approx 3.91$ K. Solid line:
$P_{c}/P_{max}=h(1-T/T_{c})^{3/2}$, where $h\approx 300$.}
\end{figure}

7. There are several differences:

1) The transition from the power at which the critical current is first
suppressed to the power at which the average critical current $I_{c}=0$ is extended
and its width is $\approx 5$ dB (Fig. 1), while in Ref. \cite{Dmitr} the width of this
transition is $\approx 0.5$ dB. The transition region depends on the absolute irradiation
power and can be large for other measurements.

2) The resistance of a single PSL $R_{0} \sim l_{E}$ most likely does not change with
an increase in power at the irradiation frequency 30 GHz, and the resistance $R$ at
the first step of the I-V characteristic of a sample with a PSL increases as a result
of the appearance of new PSL. In Ref. \cite{Dmitr} the resistance $R_{0}$ of the irradiated
sample decreased by 30 \% . It follows from the power-independence of $R_{0}$ that the
temperature dependence of $R_{0}$ is also not related to the high-frequency irradiation at
this frequency. Irradiation thus has no effect on the realization of the inelastic mechanism
of relaxation of the unbalance of the populations of the electron- and
hole-like branches of the quasiparticle spectrum.

3) In addition to a discrete increase of the resistance as a function of the
power, there are also regions where $R$ increases continuously (Fig. 2, inset). The increase
of the resistance from $R_{0}$ up to values $\approx 2R_{0}$ can be explained by a
gradual appearance of a second PSL. This process is associated with the production of
Abrikosov flux lines by an rf magnetic field. The motion of these flux lines under the
action of the measuring current gives an additional contribution to $R$. The
experimental power dependence of this contribution has the form
\begin{equation} R_{flow}=R-R_{0}=b(P/P_{max}-P_{c}/P_{max})^{1/2}\;,
\end{equation}
 where $b$ is a constant. The power $P_{c}$ at which the critical
current vanishes is most likely equal to the power at which the flux lines
start to move. The motion of the flux lines produced by the rf
electromagnetic field probably terminates with the formation of another PSL.
However, many subsequent PSLs are formed by a jump without the participation
of flux lines. Suppression of the order parameter could be another cause of
the continuous growth of $R$. In this case Ref. \cite{Gor}
\begin{equation}\delta R=R-R_{0}\sim\delta l_{E}\sim\delta\Delta/ \Delta
\sim\overline{A}^{2}_{\omega}\sim P,
\end{equation}
where $\overline{A}_{\omega}$ is the average amplitude of the magnetic vector
potential of the electromagnetic field. According to our ideas, we worked at
lower powers. Measurements with no irradiation were performed in the
interval of isothermality of PSL. Hysteresis also did not occur at low
irradiation powers, so that the continuous growth of the resistance from
$R_{0}$ up to $\approx 2R_{0}$ on the first linear section is most
likely not associated with thermal superheating.

8. In summary, phase separation, induced by an rf field with power above the
critical power $P_{c}$, on phase-slip lines was observed in wide
superconducting films. Near the critical temperature the power is
$P_{c}\sim (1-T/T_{c})^{3/2}$. As the power increases, the resistance of the
film increases, on the whole, by an amount that is a multiple of the resistance $R_{0}$
of a single PSL with no irradiation, and this increase is associated with the successive
formation of new PSLs. The resistance $R_{0}$ and therefore the time-averaged
value of the penetration depth of a longitudinal electric field do not
depend on the irradiation power at 30 GHz. There is an additional
contribution to the resistance of the sample
$\delta R\sim (P/P_{max}-P_{c}/P_{max})^{1/2}$ which could be associated with
the motion of flux lines in an rf field.

\end{document}